\documentclass[aip,jcp,reprint,twocolumn]{revtex4-1}

\usepackage{graphicx}
\usepackage{amsmath}
\usepackage{amssymb}
\usepackage{mathrsfs}

\begin{document}

\title{Determining the DNA stability parameters for the breathing dynamics
of heterogeneous DNA by stochastic optimization}

\author{Srijeeta Talukder}
\author{Pinaki Chaudhury}
\email{pinakc@rediffmail.com}
\affiliation{Department of Chemistry, University of Calcutta, 92 A P C Road,
Kolkata 700 009, India}

\author{Ralf Metzler}
\email{metz@ph.tum.de}
\affiliation{Physics Department, Technical University of Munich,
D-85747 Garching, Germany}
\affiliation{Physics Department, Tampere University of Technology,
FI-33101 Tampere, Finland}

\author{Suman K Banik}
\email{skbanik@bic.boseinst.ernet.in}
\affiliation{Department of Chemistry, Bose Institute, 93/1 A P C Road,
Kolkata 700009, India}

\date{September 30, 2011}

\begin{abstract}
We suggest that the thermodynamic stability parameters (nearest neighbor
stacking and hydrogen bonding free energies) of double-stranded DNA molecules
can be inferred reliably from time series of the size fluctuations (breathing)
of local denaturation zones (bubbles). On the basis of the reconstructed
bubble size distribution, this is achieved through stochastic optimization of
the free energies in terms of Simulated Annealing. In particular, it is shown
that even noisy time series allow the identification of the stability parameters
at remarkable accuracy. This method will be useful to obtain the DNA stacking
and hydrogen bonding free energies from single bubble breathing assays rather
than equilibrium data.
\end{abstract}


\maketitle

\section{Introduction}

The Watson-Crick double-helical form of DNA \cite{watson} is not a static
structure: even at standard salt conditions and room temperature the base
pairs may intermittently open up and expose the otherwise protected core
of the nucleotides. Such local denaturation bubbles are usually quite
short-lived, however, the propensity of double-stranded DNA towards
formation of longer-lived bubbles can be increased by elevating temperature
or lowering the salt concentration. \cite{wartell,frank,dp,Cant,Gros} In
naturally underwound circular DNA denaturation bubbles are stabilized by
partial twist release, \cite{jozef,strick} while in modern single DNA
molecule setups bubble formation may be facilitated by the exertion of
longitudinal stretching forces. \cite{pant,rief,busta,hanke,huguet} The
preferred location of bubbles is connected with the stability landscape of
the genome, as quantified by maps of stability parameters, which are
functions of the specific, underlying sequence of GC and AT base pairs.
\cite{slucia,blake,krueger,everaers,blossey,huguet} In a biological
context, bubbles correspond to so-called DNA Unwinding Elements (DUE), which
are central in processes such as gene regulation, DNA replication, and
transcription. \cite{sinden} Similarly, in higher organisms the thermodynamic
stability landscape of DNA is related to the coding versus non-coding properties
of the genome. \cite{yeramian,carlon} The denaturation of a long DNA chain from
double-strand to two separate single-strands is a physical phase transition,
whose order is determined by the magnitude of the critical exponent $c$ for
the entropy loss of a flexible polymer loop, see the discussion below.
\cite{frank,dp,fisher,wartell,kafri,hanke,Gros} The opening-closing dynamics of
denaturation bubbles can be quantified by simple nonequilibrium models based
on the gradient of the DNA stability free energy landscape.
\cite{hanke1,bicout,fogedby,kafri1,novotny}

Melting profiles of DNA can be obtained from a host of experimental techniques.
These include UV spectroscopic methods, \cite{Cant} circular dichroism,
\cite{Cant}
fluorescence resonant energy transfer measurements, \cite{Gelf} calorimetry,
\cite{Sen} or nuclear magnetic resonance, \cite{gueron} among others. Single
DNA manipulation techniques such as unzipping have recently been shown to
provide high accuracy results for the stability parameters and their salt
dependence. \cite{huguet} From the respective melting or unzipping curves the
DNA stability parameters are deduced, which in bioinformatics serve to predict
the melting profiles of arbitrary, given DNA sequences. \cite{Zuk} Up until now
the different sets of stability parameters differ considerably from each other.
\cite{slucia,blake,krueger,everaers,blossey,huguet} Alternative methods to
measure these may help to pin down optimized parameters. One way could be to
use dynamic information from bubble breathing. Indeed, by fluorescence
correlation spectroscopy the breathing dynamics of single DNA bubbles has
been monitored, producing the breathing-induced fluorescence-fluorescence
correlation function, that is pronouncedly non-exponential. \cite{altan,skb}
Given the recent progress in experimental methods, we expect that time series
of single bubble dynamics will soon become available, in which opening or
closing events of individual base pairs can be monitored. A high potential
for such time records lies in nano-channel approaches as the one reported in
Ref.~\onlinecite{jonas}, after new labeling techniques will become available
shortly.

In what follows we pursue the question whether the bubble size distribution
obtained from single breathing time
series may, in principle, be used to obtain reliable information on the DNA
stability parameters. We show that indeed by stochastic analysis methods such
as Simulated Annealing (SA) accurate estimates for the stability parameters
may be obtained for known DNA sequences.

The paper is structured as follows. We first introduce the general statistical
model of DNA base pairing, before proceeding to present the methodology of SA.
In the subsequent section we present our results, before drawing our conclusions.

\section{Statistical model for DNA denaturation}

\subsection{Thermodynamics}

The size of denaturation bubbles typically ranges from a few broken base pairs
(bps) at physiological temperature in linear, unconstrained DNA, to some 200
broken bps closer to the melting temperature of the DNA.
\cite{wartell,dp,frank,krueger,skb} Bubbles of some hundred broken base
pairs also occur in naturally underwound DNA. \cite{jozef,sinden} Following
the notation of Ref.~\onlinecite{krueger}, the stability of DNA is characterized by
the free energies $\epsilon_{hb}(\rm AT)$ and $\epsilon_{hb}(\rm GC)$ for the
Watson-Crick hydrogen bonds between complementary nucleotides (A and T, G and C,
respectively) as well as the independent stacking free energies $\epsilon_{st}$
for disrupting the stacking interactions between nearest neighbor bps. These
stacking energies depend on the nature of the two vicinal bps, as well as on
their orientation along the DNA molecule ($3^{\prime}$ to $5^{\prime}$). 
The free energies are
functions of temperature and salt concentration. Depending on the used set of
stability parameters more or less pronounced asymmetries in the stacking free
energies are observed. \cite{slucia,blake,krueger,everaers,blossey,huguet} In
addition to the hydrogen bonding and stacking free energies, there is an
additional energetic cost for initiating a bubble in the first place. Roughly
speaking, this term originates from the fact that two stacking contacts need
to be broken, while only one single broken bp yields an entropic gain. This
is either taken into consideration by the cooperativity factor $\sigma_0$, or
the so-called ring factor $\xi$, see below. \cite{REMM}

The L33B9 sequence \cite{zeng} we are analyzing in the present work is given
as follows,
\begin{equation}
\label{seq}
\begin{array}{l}
\mathtt{5'-cCGCCAGCGGCCTTTACTAAAGGCCGCTGCGCc-3'},
\end{array}
\end{equation}
where the double-strand is completed by adding the complementary single strand.
The sequence (\ref{seq}) is linear, and the high content of more stable GC bps
at the two ends ensures that these ends preferentially remain closed.
A denaturation bubble forms in the center of the chain that is rich in weaker
AT bonds. We therefore view the two extremities denoted by the lower case symbol
\texttt{c} as completely clamped. Labeling the sequence of bps by the coordinate
$x$, ranging from $x=0$ to $x=M+1$, we thus have $M=31$ internal bps, which
are allowed to open up, while the bps at $x=0$ and $x=M+1$ remain closed by
definition. In a mathematical sense, the bps at the two extremities represent
reflecting boundary conditions. Furthermore, we call $x_L$ and $x_R$ the
momentary
positions of the two closed bps embracing the denaturation bubble to the left
and right, such that the bubble size becomes $m=x_R-x_L-1$. In terms of the
Boltzmann factors for hydrogen bonding of the bp at position $x$,
\begin{equation}
\label{swth}
u_{hb}(x)=\exp\left(\frac{\epsilon_{\mathrm{hb}}(x)}{k_BT}\right),
\end{equation}
and the stacking interactions between the bps at positions $x-1$ and $x$,
\begin{equation}
\label{swts}
u_{st}(x)=\exp\left(\frac{\epsilon_{\mathrm{st}}(x)}{k_BT}\right),
\end{equation}
the bubble partition function becomes ($m\geqslant1$):
\begin{equation}
\label{pf}
\mathscr{Z}(x_L,m)=\frac{\xi'}{(1+m)^c}\prod_{x=x_L+1}^{x_L+m}u_{\mathrm{hb}}(x)
\prod_{x=x_L+1}^{x_L+m+1}u_{\mathrm{st}}(x).
\end{equation}
At $m=0$, we take $\mathscr{Z}(m=0)=1$. In Eq.~(\ref{pf}), the factor $(1+m)^{
-c}$ takes care of the entropy loss upon formation of a closed polymer loop.
For a self-avoiding chain in three dimensions, the critical exponent becomes
$c=1.76$ .\cite{fisher} Corrections of $c$ may occur due to interactions with
the rest of the chain, \cite{kafri} however, for the short DNA construct used
here, such effects are not expected to be relevant. The ring factor is $\xi
\approx10^{-3}$, \cite{krueger} and we define $\xi'=2^c\xi$. The ring factor
may be interpreted as the cooperativity parameter, divided by the Boltzmann
factor for stacking, $\xi=\sigma_0/\exp(\epsilon_{\mathrm{st}}/k_BT)$.
\cite{krueger} In principle, the ring factor depends on the position. However,
a bubble will statistically always form at the weakest link. Considering this
we have used a constant value of ring factor, $\xi$ in the present work.
With above notation, the equilibrium distribution for finding
a bubble of size $m$ and with the leftmost broken bp located at position $x+1$,
is given by
\begin{equation}
\label{eqd}
P_{\mathrm{eq}}(x_L,m)=\frac{\mathscr{Z}(x_L,m)}{\mathscr{Z}(0)+\sum_{m=1}^M
\sum_{x_L=0}^{M-m}\mathscr{Z}(x_L,m)}.
\end{equation}

\subsection{Nonequilibrium: bubble breathing}

Powered by thermal fluctuations, the bubble size becomes a random process as a
function of time. Varying stepwise by further unzipping of one bp at position
$x_L$ or $x_R$, or by zipping at $x_L+1$ and $x_R-1$, the bubble size $m$
performs a random walk along the coordinate $x$, the bubble breathing dynamics.
\cite{hanke1,fogedby,kafri1,skb,bicout,novotny} This process is described by
the master equation \cite{skb}
\begin{equation}
\label{meq}
\frac{\partial P(x_L,m,t)}{\partial t}=\mathbb{W}P(x_L,m,t),
\end{equation}
where $P(x_L,m,t)$ is the probability distribution for finding a bubble of size
$m$ with the leftmost open bp at position $x_L+1$, at time $t$. The matrix
$\mathbb{W}$ contains the transfer rates for all possible transitions in the
$(x_L,m)$ space, for details see Ref.~\onlinecite{skb}. In the long time limit, the
solution $P$ of the master equation (\ref{meq}) equilibrates to the distribution
$P_{\mathrm{eq}}$ of Eq.~(\ref{eqd}). To generate individual bubble breathing
time series for $m(t)$ and $x_L(t)$, as well as construct the distribution $P_{
\mathrm{eq}}$, one may employ the Gillespie algorithm. \cite{dtg,suman}

Following the experimental setup in Ref.~\onlinecite{altan}, one may study the
dynamics of a \emph{tagged\/} bp located at $x=x_T$. In the typical experimental
scenario fluorescence occurs if the bps in a $\delta$-neighborhood of the
fluorophore position $x_T$ are open. Measured fluorescence time series thus
correspond to the stochastic variable $I(t)$, with the properties $I(t)=1$ if
at least all bps in $(x_T-\delta, x_T+\delta)$ are open, and $I(t)=0$ otherwise. \cite{skb}
In what follows we probe whether a single bp is open or closed, i.e., we choose
$\delta=0$.

%%%%% Figure 1
\begin{figure}[t!]
\label{profile}
\includegraphics[width=1.0\columnwidth,angle=0]{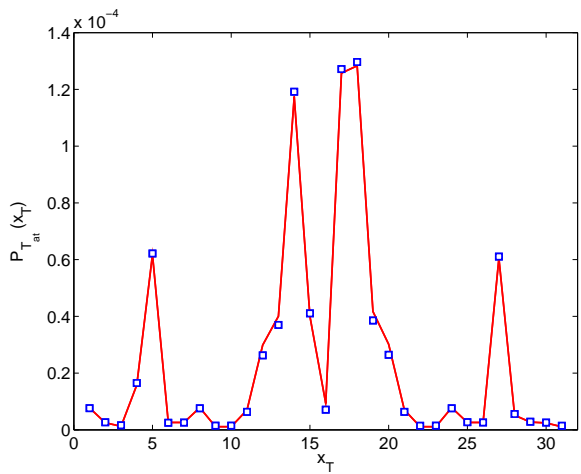}
\caption{(Color online) Theoretical probability distribution for finding a
tagged bp at position $x_T$ open (solid line), compared with the result from
the converged SA scheme (blue open squares). The underlying DNA sequence is
given in Eq.~(\ref{seq}).}
\end{figure}

\section{Stochastic optimization}

Given the probability distribution $P_{\mathrm{eq}}(m,x_L)$, constructed from
an experimental or simulations time series $m(t)$, $x_L(t)$, for a bubble in
the DNA construct under consideration: can we reliably extract the stability
parameters? Here we show that stochastic optimization is the method of choice.

Finding system parameters in a complex landscape is a generic task across
disciplines. \cite{Fogar,Pulay,Sch,Head,Wales,Bacelo,Berne} Typically, a given
problem is cast in such a manner that the seeked-for optimum corresponds to an
extremum of a functional in the complex search space. For instance, to obtain
the global minimum in a rugged potential energy surface, one starts from any
arbitrary point on this landscape and then moves on in the search space,
following certain rules, such as accepting a move if the gradient norm for the
new position decreases. This process converges to a point for which the gradient
norm is zero. To verify whether this point is a minimum, one needs to check if
the eigenvalues of the Hessian matrix at that point are all positive.
A completely deterministic optimization procedure such as this minimization of
the gradient norm, however, will generally fail to determine the global minimum
if the search space features multiple minima. Once a local minimum is found,
the deterministic search method will simply terminate. Such a misguidance is
avoided by true global optimizers, whose search is not solely driven by a
gradient. In particular, stochastic optimization techniques turn out to be very
successful. Originally proposed by Kirkpatrick and coworkers to solve the
traveling salesman problem, \cite{ksk1,ksk2} SA
represents such a true global
optimizer, and has been applied to a broad range of problems across disciplines,
see, for instance, Refs.~\onlinecite{Car,Nandy,PDutta,Ming,Zuck,Aarts,Noller,Kalb,Franc}. 
In SA, the search space is initially sampled at a high temperature
($T_{\mathrm{at}}$). The associated thermal fluctuations at a suitable
value of $T_{\mathrm{at}}$ will lift the optimizer out of local minima such
that the search may continue towards increasingly deeper minima. Once the
temperature becomes sufficiently small and/or the search is carried out over
a sufficient time span, the entire search space is probed. Due to this ergodic
property the global minimum is indeed found unequivocally. \cite{ksk2}

Typically, an SA analysis is started at a sufficiently high temperature. This
makes nearly all moves acceptable, as the criterion for accepting or rejecting
a move is determined by the Metropolis criterion. In our case, the associated
cost function, which is being minimized, is the sum of the squares of the
difference of the occupation probabilities at the various positions,
\begin{equation}
\label{eq1}
{\rm cost}_i=\sum_{i=1}^{M}(P_{\mathrm{eq}}(x_i)-P_{T_{\mathrm{at}}}(x_i))^2,
\end{equation}
where $P_{T_{\mathrm{at}}}(x)$ denotes the distribution at position $x$ found
in the current SA step, when the simulation temperature is $T_{\mathrm{at}}$.
If, on going from one SA step ($i$) to the next ($i+1$) the magnitude of the
cost function decreases, we at once accept that move. If it increases, we do
not discard the move rightout. Instead, we subject it to the Metropolis test
\cite{metrop}: if the quantity $\Delta={\rm cost}_i-{\rm cost}_{i-1}$ has a
positive value, the probability for accepting the move is determined by the
function
\begin{equation}
\label{eq2}
F=\exp\left(-\frac{\Delta}{T_{\mathrm{at}}}\right).
\end{equation}
For positive $\Delta$, $F$ is always between 0 and 1. For each evaluation
of $F$, we invoke a random number $\mathrm{rand}$ between 0 and 1. If $F>
\mathrm{rand}$, we accept the move. If not, the move is rejected. Thus, at
very high $T_{\mathrm{at}}$, $F$ will be close to 1 and most moves will be
accepted, such that a greater region of the search space will be sampled.
As the simulation proceeds, $T_{\mathrm{at}}$ is decreased by the 
\emph{annealing schedule}. 
Once the correct path towards the global minimum is followed, we
need not search the entire space and concentrate on a small region, which will
guide us specifically to the global minimum. That is, as $T_{\mathrm{at}}$ is
lowered, a decreasing number of moves pass the Metropolis test. Ultimately, in
our problem we recover the stability parameters from the SA analysis.

 In SA, the crucial factor which determines the success of optimization is the
 annealing schedule, which is basically the rate at which the simulation temperature
 is decreased in successive annealing steps. In the present study we have kept
 the initial temperature at 1000. The rate of cooling was kept at $10\%$ of the
 value of the present step. We have also ensured that after every 30 SA steps,
 the system is re-heated to the initial starting value, i.e., the simulation temperature
 is forcibly increased to a higher value. This is done to remove any possibility of
 being trapped in a local minimum (coming out of which will be difficult if the
 simulation temperature is low). In successive SA steps, along with the temperature,
 the individual stability parameters are changed by the following strategy.
 If $u$ is a parameter chosen for change in SA, it is updated by the rule:
$u^{'} = u+u \times (-1)^{n} \times \delta \times r_n$,
where $n$ is a random integer, $\delta$ is the amplitude of allowed change (kept
at $0.01$), and $r_n$ is a random number between $0$ and $1$. The new $u{'}$
(changed stability parameter) is used to generate the updated distribution
profile. The magnitudes of the different optimization parameters are collected in
Table~\ref{mag}.

%%%%% Figure 2
\begin{figure}[!b]
\includegraphics[width=1.0\columnwidth,angle=0]{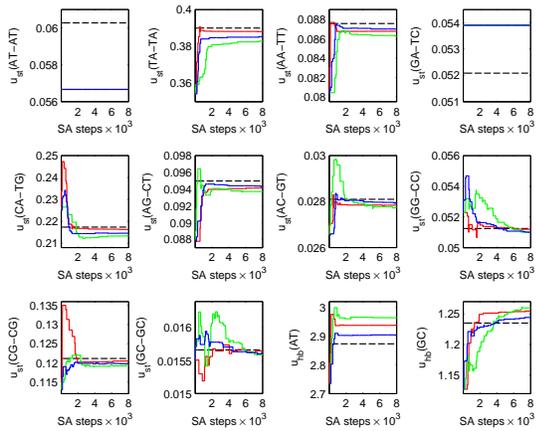}
\caption{(Color online) Evolution of the free energy parameters of hydrogen
bonding and base stacking as function of SA steps (full lines) from three
separate SA runs. The black dashed horizontal lines represent the expected
experimental values taken from Ref.~\onlinecite{krueger}, towards which
convergence is expected to occur. Note the different scales on the vertical
axes. The values for two pairs of bps, AT-AT and GA-TC, do not change in the
SA procedure; these two pairs do not occur in the underlying sequence
(\ref{seq}) and are thus not subject to the SA optimization criteria, i.e.,
they do not converge.}
\label{prog}
\end{figure}

%%%%% Table I
\begin{table}
\caption{Magnitude of optimization parameters used in SA.}
\begin{ruledtabular}
\begin{tabular}{lcc}
 Parameter & Magnitude \\
\hline
 Annealing Schedule   & $10\%$   \\
 Initial Simulation Temperature  &  1000 \\
 Magnitude of Change $\delta$  &   0.01   \\
\end{tabular}
\end{ruledtabular}
\label{mag}
\end{table}

\section{Results and Discussion}

In a first step, the equilibrium distribution for a tagged bp at location
$x_T$ in the DNA sequence (\ref{seq}) was determined from the theoretical
stability parameters from Ref.~\onlinecite{krueger}. SA was then employed
for successive convergence of $P_{T_{\mathrm{at}}}$ to this theoretical
distribution through variation of the 12 independent free energy parameters
(compare Table~\ref{tab}), by minimizing the cost function. The SA analysis
was terminated once the value of the cost function becomes smaller than
$10^{-4}$. Fig.~1 shows the quite accurate convergence of the SA
scheme in terms of the equilibrium distribution.

To visualize the progress of the SA procedure, we display in Fig.~\ref{prog}
the progress of the approximation of the twelve DNA stability parameters of
hydrogen bonding and base stacking (compare also Table~\ref{tab}) for 8000
SA steps, for three separate SA runs starting with different initial simulation
temperatures. For each simulation the initial free energy values are chosen
via random perturbation of the experimental $u$ values, \cite{krueger}
following our SA strategy.
In all cases the convergence is quite accurate. Two parameters
do not change during the SA scheme, these correspond to the two pairs of bps,
that do not occur in the employed sequence (\ref{seq}). To be sure that the
search proceeds without being held up in local basins, the annealing
temperature was raised after every 30 SA steps and then allowed to follow the
usual annealing schedule. The sudden jumps in the profile are a result of this
effort. At an abruptly elevated temperature, newer moves start to get accepted
and hence the zigzag pattern.

%%%%% Figure 3
\begin{figure}[!b]
\label{zcmb}
\includegraphics[width=1.0\columnwidth,angle=0]{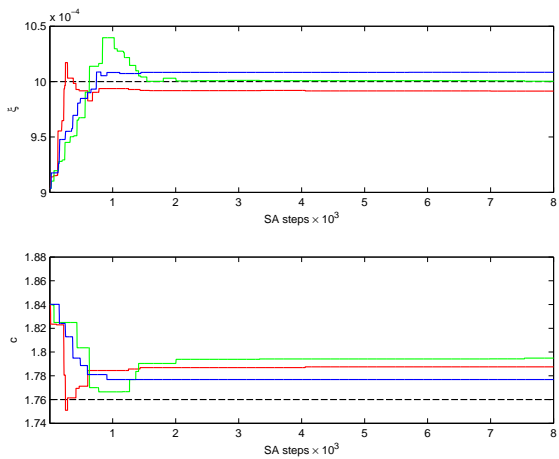}
\caption{(Color online) Evolution of ring factor $\xi$ and critical exponent $c$
from three different SA runs. The black dashed horizontal lines represent the 
expected literature value towards which convergence is expected to occur.}
\end{figure}

In terms of the free energy values for hydrogen bonding and base stacking, the
average results from 1000 SA runs are shown in Table~\ref{tab}. We also
indicate which combinations of nearest neighbor pairs actually occur in the
underlying sequence (\ref{seq}). The convergence of the SA algorithm in all
cases is quite remarkable. In addition to the free energy parameter we also
optimized the loop exponent $c$ and the ring factor $\xi$. The resultant simulation
profiles (Fig.~3) show a good convergence towards theoretical values.

In typical experimental data the distribution of the bubble opening probability
will be noisy, due to finite sampling and measurement errors. To check if our
SA algorithm is robust against such noise we randomly perturbed the
theoretically expected equilibrium distribution by a gaussian random processes
with amplitude and width being the $P_{eq}$ and 10$\%$ of $P_{eq}$,
respectively.
Fig.~4 shows how this noisy data was quickly smoothened out to
reach the theoretical distribution profile. We show snapshots of the
process for different SA steps. In each figure, the original noisy
data, the equilibrium distribution profile and the evolving
profile at the particular SA step are shown. At 1500 SA steps, the
noisy data completely matches with the equilibrium distribution.

%%%%% Figure 4
\begin{figure}
\label{fig4}
\includegraphics[width=1.0\columnwidth,angle=0]{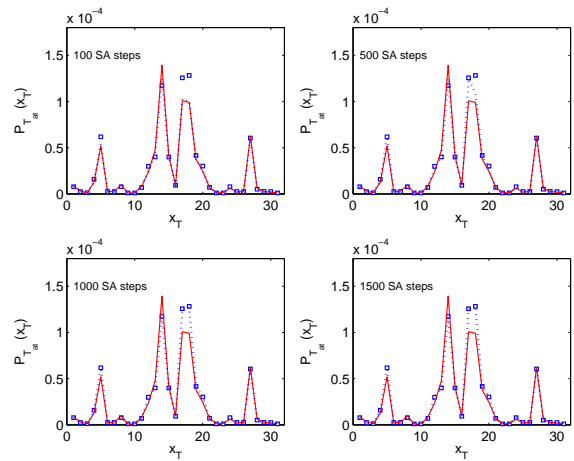}
\caption{(color online) Plot of $P_{T_{\mathrm{at}}}(x_T)$ against $x_T$ at
various SA steps. In each panel the red solid lines represents the original
noisy data, and the blue dashed line is the output of SA runs. The blue open
squares stand for the theoretical distribution $P_{\mathrm{eq}}(x_T)$. The
plot for 1,500
SA steps already matches quite well the expected distribution $P_{eq}(x_T)$.}
\end{figure}

%%%%% Table II
\begin{table}
\caption{Comparison of experimental \cite{krueger} and simulated free energy
data. Each simulation data is a mean of 1000 different SA outputs. The
rightmost column shows the presence ($\surd$) or absence ($\times$) of
particular free energies in sequence (\ref{seq}). Units of free energies
($\epsilon_{st}$ and $\epsilon_{hb}$) reported here are kcal/mol. The
last two rows of the table gives a comparison of the ring factor $\xi$ and
critical exponent $c$.
}
\begin{ruledtabular}
\begin{tabular}{lccc}
& Experimental & SA results & \\
%& (kcal/mol) & (kcal/mol) &  \\
\hline
$\epsilon_{st}$ (AT-AT)   &   -1.729409   &  -1.767474   & $\times$ \\
$\epsilon_{st}$ (TA-TA)   &  -0.579800    &  -0.588968   & $\surd$ \\
$\epsilon_{st}$ (AA-TT)   &   -1.499484   & -1.510239    & $\surd$ \\
$\epsilon_{st}$ (GA-TC)  &   -1.819371   & -1.798201    & $\times$ \\
$\epsilon_{st}$ (CA-TG)  &    -0.939677  & -0.922743    & $\surd$ \\
$\epsilon_{st}$ (AG-CT)  &   -1.455363   & -1.462615    & $\surd$ \\
$\epsilon_{st}$ (AC-GT)  &    -2.199241  & -2.175124    & $\surd$ \\
$\epsilon_{st}$ (GG-CC) &    -1.829370  & -1.801741    & $\surd$ \\
$\epsilon_{st}$ (CG-CG) &    -1.299554  & -1.318516    & $\surd$ \\ 
$\epsilon_{st}$ (GC-GC) &    -2.559130  & -2.549840    & $\surd$ \\
$\epsilon_{hb}$ (AT)  &     0.649775    &    0.651781   & $\surd$ \\
$\epsilon_{hb}$ (GC) &      0.129955   &    0.113848   & $\surd$ \\
\hline
$\xi$         &    0.001 &  0.001034062  & \\    
c &     1.76         &    1.758298    & \\
\end{tabular}
\end{ruledtabular}
\label{tab}
\end{table}

\section{Conclusion}

Generalising our previous approach, \cite{obdyn} we here demonstrate the
outstanding ability of stochastic optimization to determine the stability
parameters of double-stranded DNA from time series of the breathing dynamics
of individual bps. Even for a short DNA sequence such as L33B9 [Eq.~(\ref{seq})]
with only 31 internal bps, the convergence of the chosen SA
scheme to all present base stacking and hydrogen bonding free energies is
recovered with appreciable accuracy. Even when the input data are perturbed
randomly, mimicking noisy experimental or simulations data, the stochastic
optimization technique works successfully.

Optimization based on the bubble distribution $P_{\mathrm{eq}}(x)$ is not the
only way
to extract the DNA stability parameters. For instance, one might use average
values for the zipping and unzipping rates of individual bps and relate their
ratio to the underlying free energy difference. Alternatively, once from high
throughput fluorescence correlation experiments an accurate result for the
fluorescence autocorrelation function becomes available, one might use this
function as basis for the optimization. In principle, one might also modify
our approach to analyse data from DNA unzipping. This, however, requires
detailed knowledge on the change of the stacking and hydrogen free energies
upon stretching of the DNA strands.

In general, it may be worthwhile to also explore the possibility to apply
other techniques such as the genetic algorithm, \cite{Gold} parallel tempering, 
\cite{Earl} or ant colony optimization, \cite{Dorigo,Bon} and to compare
these methods.

\begin{acknowledgments}
ST acknowledges the financial support form UGC, New Delhi, for
granting a Junior Research Fellowship [UGC/800/Jr. Fellow (SC)].
PC wishes to thank The Centre for Research on Nano Science and Nano Technology,
University of Calcutta for a research grant [Conv/002/Nano RAC (2008)].
RM acknowledges funding through the Academy of Finland's FiDiPro scheme.
SKB acknowledges support from Bose Institute through a initial start up fund.
\end{acknowledgments}


\begin{thebibliography}{99}

\bibitem{watson} J. D. Watson and F. H. C. Crick, Nature \textbf{171}, 737
(1953); R. E. Franklin and R. G. Gosling, Nature \textbf{171}, 740 (1953);
F. Crick, Nature \textbf{227}, 561 (1970).

\bibitem{frank} M. D. Frank-Kamenetskii, Phys. Rep. \textbf{288}, 13 (1997).

\bibitem{wartell} R. M. Wartell and A. S. Benight, Phys. Rep. \textbf{126},
67 (1985).

\bibitem{Gros} A. Y. Grosberg and A. R. Khokhlov,
\emph{Statistical Physics of Macromolecules}
(AIP Press, New York, 1994).

\bibitem{dp} D. Poland and H. A. Scheraga,
\emph{Theory of Helix-Coil Transitions in Biopolymers}
(Academic Press, New York, 1970).

\bibitem{Cant} C. R. Cantor and P. R. Schimmel,
\emph{Biophysical Chemistry}
(W H Freeman, New York, 1980).

\bibitem{strick} T. R. Strick, V. Croquette, and D. Bensimon, Nature
{\bf 404}, 901 (2000).

\bibitem{jozef} J.-H. Jeon, J. Adamczik, G. Dietler, and R. Metzler,
Phys. Rev. Lett. \textbf{105}, 208101 (2010).

\bibitem{pant} M. C. Williams, J. R. Wenner, I. Rouzina, and V. A. Bloomfield,
Biophys. J. \textbf{80}, 874 (2001); K. R. Chaurasiya, T. Paramanathan, M. J.
McCauley, and M. C. Williams, Phys. Life Rev. \textbf{7}, 299 (2010).

\bibitem{rief} M. Rief, H. Clausen-Schaumann, and H. E. Gaub, Nature Struct.
Biol. \textbf{4}, 153 (1997);
H. Clausen-Schaumann, M. Rief, C. Tolksdorf, and H. E. Gaub,
Biophys. J. \textbf{78}, 1997 (2000).

\bibitem{busta} S. B. Smith, Y. J. Cui, and C. Bustamante, Science \textbf{271},
795 (1996).

\bibitem{hanke} A. Hanke, M. G. Ochoa, and R. Metzler, Phys. Rev. Lett.
\textbf{100}, 018106 (2008).

\bibitem{huguet} J. M. Huguet, C. V. Bizarro, N. Forns, S. B. Smith, C.
Bustamante, and F. Ritort, Proc. Natl. Acad. Sci. USA \textbf{107}, 15431 (2010).

\bibitem{krueger} A. Krueger, E. Protozanova, and M. D. Frank-Kamenetskii,
Biophys. J. \textbf{90}, 3091 (2006). The parametrisation of the DNA stability
free energies in this work makes it possible to distinguish the stacking and
the hydrogen bonding free energies.

\bibitem{blake} R. D. Blake, J. W. Bizzaro, J. D. Blake, G. R. Day, S. G.
Delcourt, J. Knowles, K. A. Marx, and J. SantaLucia, Jr.,
Bioinformatics \textbf{15}, 370 (1999).

\bibitem{slucia} J. SantaLucia, Jr.,
Proc. Natl. Acad. Sci. U.S.A. \textbf{95}, 1460 (1998).

\bibitem{everaers} D. Jost and R. Everaers, Biophys. J. \textbf{96}, 1056
(2009).

\bibitem{blossey} R. Blossey and E. Carlon, Phys. Rev. E \textbf{68}, 061911
(2003).

\bibitem{sinden} R. R. Sinden,  \emph{DNA Structure and Function} (Academic
Press, San Diego, CA, 1994).

\bibitem{carlon} E. Carlon, M. L. Malki, and R. Blossey, Phys. Rev. Lett.
\textbf{94}, 178101 (2005).

\bibitem{yeramian} E. Yeramian, Gene \textbf{255}, 139 (2000).

\bibitem{fisher} M. E. Fisher, J. Chem. Phys. \textbf{44}, 616 (1966).

\bibitem{kafri} Y. Kafri, D. Mukamel, and L. Peliti, Phys. Rev. Lett.
\textbf{85}, 4988 (2000); Euro. Phys. J. B \textbf{27}, 132 (2002).

\bibitem{hanke1} A. Hanke and R. Metzler, J. Phys. A \textbf{36}, L473 (2003).

\bibitem{bicout} D. Bicout and E. Kats, Phys. Rev. E \textbf{70}, 010902(R)
(2004).

\bibitem{kafri1} A. Bar, Y. Kafri, and D. Mukamel, Phys. Rev. Lett.
\textbf{98}, 038103 (2007).

\bibitem{novotny} T. Novotn{\'y}, J. N. Pedersen, T. Ambj{\"o}rnsson, M. S. Hansen,
and R. Metzler, Europhys. Lett. \textbf{77}, 48001 (2007); J. N. Pedersen, M. S.
Hansen, T. Novotn{\'y}, T. Ambj{\"o}rnsson, and R. Metzler, J. Chem. Phys.
\textbf{130}, 164117 (2009).

\bibitem{fogedby} H. C. Fogedby and R. Metzler, Phys. Rev. Lett. \textbf{98},
070601 (2007); Phys. Rev. E \textbf{76}, 061915 (2007).

\bibitem{Gelf} C. A. Gelfand, G. E. Plum, S. Mielewczyk, D. P. Remeta and K. J. Breslauer,
Proc. Natl. Acad. Sci. U. S. A. \textbf{96}, 6113 (1999)

\bibitem{Sen} M. M. Senior, R. A. Jones K. J. Breslauer,
Proc. Natl. Acad. Sci. U. S. A. \textbf{85}, 6242 (1988)

\bibitem{gueron} M. G{\'e}ron, M. Kochoyan, and J.-L. Leroy, Nature \textbf{328},
89 (1987).

\bibitem{Zuk} M. Zuker,
Nucl. Acids Res. \textbf{31}, 3406 (2003).

\bibitem{altan} G. Altan-Bonnet, A. Libchaber, and O. Krichevsky, Phys. Rev.
Lett. \textbf{90}, 138101 (2003).

\bibitem{skb} T. Ambj\"ornsson, S. K. Banik, O. Krichevsky and R. Metzler,
Phys. Rev. Lett. \textbf{97}, 128105 (2006); Biophys. J. \textbf{92}, 2674
(2007); T. Ambj\"ornsson, S. K. Banik, M. A. Lomholt and R. Metzler,
Phys. Rev. E \textbf{75}, 021908 (2007).

\bibitem{jonas} W. Reisner, N. B. Larsen, A. Silahtaroglu, A. Kristensen,
N.Tommerup, J. O. Tegenfeldt, and H. Flyvbjerg, Proc. Natl. Acad. Sci. USA
\textbf{107}, 13294 (2010).

\bibitem{REMM} Note that the bubble will on average open at the weakest
base pair-base pair couple. The ring factor $\xi$ or cooperativity constant
$\sigma_0$ is therefore independent of the index $x$.

\bibitem{zeng} Y. Zeng, A Montrichok, and G. Zocchi,
J. Mol. Biol. \textbf{339}, 67 (2004).

\bibitem{dtg} D. T. Gillespie,
J. Comput. Phys. \textbf{22}, 403 (1976);
J. Phys. Chem. \textbf{81}, 2340 (1977).

\bibitem{suman} S. K. Banik, T. Ambj{\"o}rnsson, and R. Metzler, Europhys.
Lett. \textbf{71}, 852 (2005).

\bibitem{Fogar} G. Fogarsi and P. Pulay,
Ann. Rev. Phys. Chem. \textbf{35}, 191 (1984).

\bibitem{Pulay} P. Pulay and J. Simons, eds.,
\emph{Geometrical Derivatives of Energy Surfaces and Molecular Properties}
 (Reidel, Dordecht, 1986).

\bibitem{Sch} H. B. Schlegel,
Adv. Chem. Phys. \textbf{67}, 249 (1987).

\bibitem{Head} J. D. Head, B. Weiner and M. C. Zerner,
Int. J. Quant. Chem. \textbf{33}, 177 (1988).

\bibitem{Wales} M. C. Prentiss, D. J. Wales and P. G. Wolynes,
J. Chem. Phys. \textbf{128}, 225106 (2008).

\bibitem{Bacelo} D. E. Bacelo and  S. E. Fioressi,
J. Chem. Phys. \textbf{119}, 11695 (2003).

\bibitem{Berne} P. Liu and B. J. Berne,
J. Chem. Phys. \textbf{118}, 2999 (2003).

\bibitem{ksk1} K. S. Kirkpatrick, C. D. Gelatt, and M. P. Vecchi,
Science \textbf{220}, 671 (1983).

\bibitem{ksk2} K. S. Kirkpatrick,
J. Stat. Phys. \textbf{34}, 975 (1984).

\bibitem{Car} R. Car and M. Parinello,
Phys. Rev. Lett. \textbf{55}, 2471 (1985).

\bibitem{Nandy} S. Nandy, P. Chaudhury, R. Sharma and S. P. Bhattacharyya,
J. Theor. Comp. Chem. \textbf{7}, 977 (2008).

\bibitem{PDutta} P. Dutta, D. Mazumdar and S. P. Bhattacharyya,
Chem. Phys. Lett. \textbf{181}, 288 (1991).

\bibitem{Ming} J. Mingjun and T. Huanwen,
Chaos Solitons Fractals, \textbf{21}, 933 (2004).

\bibitem{Zuck} E. Lyman and D. M. Zuckerman,
J. Chem. Phys. \textbf{127}, 065101 (2007).

\bibitem{Aarts} P. J. M. van Laarhoven and E. H. L. Aarts,
\emph{Simulated Amnnealing Theory and Applications}
(Kluwer Academic, Dordecht, Holland 1987).

\bibitem{Noller} A. Korostelev, M. Laurberg and H. F. Noller,
Proc. Natl. Acad. Sci. U. S. A. \textbf{106}, 18195 (2009).

\bibitem{Kalb} A. Moglich, D. Weinfurtner, T. Maurer, N. Gronwald 
and H. R.Kalbitzer,
BMC Bioinformatics \textbf{6}, 91 (2005).

\bibitem{Franc} D. Francois,
 Ann. Appl. Prob. \textbf{12}, 248 (2002).

\bibitem{metrop} N. Metropolis, A. W. Rosenbluth, M. N. Rosenbluth,
A. H. Teller and E. Teller,
J. Chem. Phys. \textbf{21}, 1087 (1953).

\bibitem{obdyn} P. Chaudhury, R. Metzler, and S. K. Banik,
J. Phys. A \textbf{42}, 335101 (2009).

\bibitem{Gold} D. E. Goldberg,
\emph{Genetic Algorithms in Search, Optimization and Machine Learning}
(Addison Wesley, Rading, MA, 1989).

\bibitem{Earl} D. J. Earl and M. W. Deem,
Phys. Chem. Chem. Phys. \textbf{7}, 3910 (2005).

\bibitem{Dorigo} M. Dorigo, V. Maniezzo and A. Colorni,
IEEE Transactions on Systems, Man and Cybernetics - Part - B: CYBERNETICS
\textbf{26}, 29 (1996).

\bibitem{Bon} E. Bonabeau, M. Dorigo, and G. Theraulaz,
Nature \textbf{406}, 39 (2000).

\end{thebibliography}
\end{document}